%% file: ver2ForArxiv.tex
\documentclass[pre,twocolumn]{revtex4-2}
\usepackage{amsmath}
\usepackage{amssymb}
\usepackage{graphicx}
\usepackage{braket}
\usepackage{stmaryrd}
\usepackage{hyperref}
\usepackage{color}
\usepackage{dsfont}
\usepackage{transparent}
\usepackage{bm}
\usepackage{enumitem}

\hypersetup{
    colorlinks,
    citecolor=blue,
    linkcolor=blue,
    urlcolor=blue
}

\input{setting}

\begin{document}


\title{
Thermodynamic Uncertainty Relation for Generalized Time-Reversal Observables
}

\author{Tetta Indo}
\email{indo@biom.t.u-tokyo.ac.jp}
\affiliation{Department of Information and Communication Engineering, Graduate
School of Information Science and Technology, The University of Tokyo,
Tokyo 113-8656, Japan}

\author{Yoshihiko Hasegawa}
\email{hasegawa@biom.t.u-tokyo.ac.jp}
\affiliation{Department of Information and Communication Engineering, Graduate
School of Information Science and Technology, The University of Tokyo,
Tokyo 113-8656, Japan}

\date{\today}
\begin{abstract}
Time-reversal symmetry plays an essential role in the thermodynamic uncertainty relations, which bound the fluctuations of observables in terms of the associated dissipation.
In fact, thermodynamic uncertainty relations are typically derived under the assumption that the observable of interest is antisymmetric under time reversal.
This also suggests that existing thermodynamic uncertainty relations are restricted to a limited class of observables.
In this paper, we mitigate this restriction by introducing a new class of observables 
that do not exhibit the exact antisymmetry but change the sign under time reversal.
We call it generalized time reversal and derive a broadly applicable thermodynamic uncertainty relation for observables with this condition.
The generalization is achieved by direct statistical arguments on the observable distributions and holds for both deterministic and stochastic dynamics.  
We demonstrate the derived thermodynamic uncertainty relation with observables subjected to rectifications, showing that the precision of generalized observables remains expressible in terms of the dissipative cost. 
The result extends the scope of thermodynamic uncertainty relations beyond the reach of previous frameworks relying on the exact antisymmetry.

\end{abstract}
\maketitle

\section{Introduction}
Over the past few decades, our understanding of non-equilibrium systems has been reshaped by the discovery of fluctuation theorems (FTs) \cite{jarzynski1997,crooks1999,
gallavotti1995,esposito2010}, which provide exact relations that govern the statistical properties of thermodynamic quantities like work, heat, and entropy production. These theorems are a direct consequence of the time-reversal symmetry of the underlying microscopic dynamics and represent a refinement of the second law of thermodynamics \cite{seifert2012}. More recently, an equally powerful set of principles has emerged in the form of thermodynamic uncertainty relations (TURs) \cite{barato2015thermodynamic,gingrich2016dissipation,
horowitz2020thermodynamic}. In its initial formulation, the TURs established a fundamental trade-off between the precision of a thermodynamic current and the associated dissipative cost, stating that the signal-to-noise ratio is bounded by the total entropy production \cite{barato2015thermodynamic,
gingrich2016dissipation}.
These bounds have been derived for a variety of stochastic and deterministic dynamics, from biomolecular machines to electronic transport \cite{barato2015thermodynamic,manzano2018thermodynamics}, and have proven instrumental in quantifying performance limits of small-scale engines and information processors \cite{landi2021non}.

Recent research has established a connection between FTs and TURs, enabling direct derivation of a variety of TURs from the underlying mathematical framework of FTs \cite{barato2015thermodynamic,
gingrich2016dissipation,horowitz2020thermodynamic,hasegawa2019ftur}.
This intricate relation was made clear following the discovery of a new type of FTs.
One such result is a stronger form of the detailed fluctuation theorem (DFT) for joint probability distributions, which revealed that the statistics of fluctuating observables are intrinsically correlated with each other \cite{harris2007,esposito2010,merhav2010,hasegawa2019ftur}. 
This connection provided the basis for the direct derivation of TURs naturally from the joint fluctuations.
Such correlations are typically characterized by a joint probability distribution of the measured observable and the associated entropy production. 
Formally, it is formulated by integrating the joint distribution over the space of all possible system trajectories (path), the sequence of states it evolved through~\cite{hasegawa2019ftur}.

However, the indispensable yet nontrivial basis for these arguments is the antisymmetry of the subject observable under time reversal, expressed as \mbox{$\phi(\Gamma^\dagger)=-\phi(\Gamma)$}. 
This parity plays a crucial role in quantitatively relating observables measured along a
time-reversed trajectory to their forward counterparts.
Nevertheless, this exact premise, often implicitly assumed in the literature, has also become a major constraint~\cite{horowitz2020thermodynamic,Liu2023universal}, restricting the applicability of the TURs to a specific class of physical quantities.
One approach to address this limitation is to replace dissipation with a time-symmetric cost \cite{diterlizzi2019kur,vo2022unified}. 
Dynamical activity, defined as the rate of microscopic transitions, comes with such a symmetry and captures the kinetic contributions to precision in a way analogous to TUR.
This so-called kinetic uncertainty relation (KUR) is applicable to generic observables because of the invariance of the dynamical activity under time reversal \cite{diterlizzi2019kur,hiura2021}. 
Nonetheless, its physical interpretation is less direct than that of entropy production, as each jump does not necessarily contribute to an emergent event of physical relevance, and the notion of activity is not well-defined in continuous-state systems~\cite{diterlizzi2019kur,monnai2024}.

In this paper, we derive a saturable TUR for the class of observables that do not necessarily exhibit the exact antisymmetry under time reversal.
The new class of observables is defined with a minimal parity requirement: \mbox{$\phi(\Gamma^\dagger)\phi(\Gamma)\leq0$} [cf. Eq.~\eqref{a}], which we call \textit{generalized time reversal}.
We derive our TUR by finding the joint probability distribution that gives the minimum possible scaled variance, given the relaxed restriction on the observable.
This is in contrast to the formerly known TURs, which are valid only when a strong joint symmetry exists among concurrently measured observables.
Notably, the derived bound turns out to be saturable and thus sharp.
Moreover, our derivation uses a concentration inequality. 
In the study of KURs, the concentration inequality known as the Petrov inequality \cite{PETROV20072703} has been used \cite{Hasegawa2024TCI}, whereas for TURs, the use of Cantelli's inequality and the Vysochanskij--Petunin inequality appears to be new.

We apply our TUR to a three-level quantum thermal machine \cite{Scovil1959,GevaKosloff1994,BoukobzaTannor2006,Alicki1979} autonomously absorbing heat from the environment.
It is shown that this one-way heat transfer is captured within the class of generalized time-reversal, with its precision bounded by the derived TUR.
We also show that our framework extends to other statistical inequalities and to the corresponding observables carrying additional distributional information.
This, in turn, allows us to derive a stronger TUR, and we test it in a diffusion-channel model. 
The result shows that existing TURs are violated by the generalized observables, whereas our bound remains valid.
Most notably, our bound stays finite and does not diverge in the near-equilibrium limit where the mean entropy production approaches zero.
This contrasts with conventional TURs, which can lose their physical interpretability in that regime \cite{barato2015thermodynamic,gingrich2016dissipation,horowitz2020thermodynamic}. 
In other words, our TUR provides a practical limit on the achievable precision of observables that existing bounds have been unable to capture.
\section{Methods}
Throughout the paper, we set the Boltzmann constant to unity.
We consider a general system driven out of equilibrium.
A single realization of the system's time evolution is represented by a microscopic trajectory $\Gamma:=[x(t)]_{t=0}^{t=T}$, where $x(t)$ denotes the system's configuration at time $t$.
Under suitable assumptions such as the local detailed balance condition in Markov processes~\cite{Seifert2005integralft}, the total entropy production $\sigma$ is defined along trajectories: $\sigma[\Gamma]=\ln [\mathcal{P}[\Gamma]/\mathcal{P}^{\dagger}[\Gamma^{\dagger}]],$
where the path-probability functional $\mathcal{P}^\dagger$ is of the time-reversed process.
Then, the second law of thermodynamics is recovered: $\langle\sigma\rangle\geq0$.
In this work, we focus on the situations where the forward and reverse evolutions are observationally indistinguishable.
One example is a steady-state system with no external protocol,
where the initial distribution is stationary and satisfies
\(P(x,0)=P(x,T)\) \cite{Spinney2013pedagogical}.
In such situations, the reverse process maps the system back to the time-reversed counterpart of the forward initial distribution.
Since the processes are identical, $\mathcal{P} = \mathcal{P}^\dagger$ holds, and the entropy-production definition  simplifies to $\sigma[\Gamma]=\ln [\mathcal{P}[\Gamma]/\mathcal{P}[\Gamma^\dagger]].$
From this definition, we can see that each time-reversal pair of trajectories has the trajectory-level fluctuation relation of:
\begin{equation}
    \mathcal{P}[\Gamma]=e^{\sigma[\Gamma]}\,\mathcal{P}[\Gamma^\dagger]
    .
    \label{core}
\end{equation}
Integrating over all trajectories with a given entropy-production value $\sigma$ yields a detailed fluctuation theorem (DFT) \cite{Seifert2005integralft}:
\(
  {p(\sigma)}/{p(-\sigma)} = e^{\sigma}.
\)
Finally, one can also show the following relation of entropy production,
\begin{equation}
    \sigma(\Gamma^\dagger)=-\sigma(\Gamma).   
    \label{sigmahanten}
\end{equation}

Now, we consider an observable $\phi(\Gamma)$, measured with the entropy production $\sigma(\Gamma)$ along a trajectory $\Gamma$.
To exploit the DFT in deriving TURs, a strong constraint on $\phi(\Gamma)$ is typically imposed:
\begin{equation}
  \phi(\Gamma^\dagger)=-\phi(\Gamma).
  \label{typicallyimposed}
\end{equation}
Note that, for current-type observables, this property follows directly from its definition.
With Eq.~\eqref{typicallyimposed}, one gets the strong DFT for joint distributions integrated over the space of all possible system trajectories \cite{hasegawa2019ftur}.
This integration is carried out by mapping each forward trajectory to its time-reversed counterpart, using the invariance of the path-integral measure. \cite{hasegawa2019ftur,Seifert2005integralft,Spinney2013pedagogical}.
However, such a mapping is valid only when $\phi(\Gamma^\dagger)$ is automatically translated into $-\phi(\Gamma)$.

Our main interest in the present work is in relaxing this constraint on the observable $\phi$.
Here, we introduce the generalized time-reversal property:
\begin{equation}
  \phi(\Gamma^\dagger)\phi(\Gamma)\leq0.\label{a}
\end{equation}
While condition \eqref{a} is strictly weaker than the odd‐parity requirement of Eq.~\eqref{typicallyimposed}, it still enforces the key feature of the sign-reversal structure akin to entropy production: $\phi(\Gamma)$ and $\phi(\Gamma^\dagger)$ cannot share the same sign.
This structure establishes constraints between $\phi$ and entropy production.
As we discuss later, this relaxation allows us to capture a broad class of physically relevant observables.

\section{RESULTS}

Our main result is the following TUR for generalized time-reversal observables, which is expressed in terms of the inverse function $f = x^{-1}$ of $x(z)=z\tanh\!\left( z/2\right)$:
\begin{equation}
\frac{\mathrm{Var}[\phi]}{\langle\phi\rangle^2}
\;\ge\;
e^{-f(\langle\sigma\rangle)}
.
\label{res}
\end{equation}
Essentially, the inverse function $f$ determines the entropy production at which the minimum relative variance is achieved within the generalized time-reversal class.

A key observation is that in contrast to existing TURs~\cite{hasegawa2019ftur,timpanaro2019tur_exchange,barato2015thermodynamic}, our bound stays finite even in the equilibrium limit.
For conventional antisymmetric observables, the relative variance typically diverges as \(\langle\sigma\rangle\to0\) since the equilibrium ensemble gives equal probability to $\Gamma$ and its reverse $\Gamma^\dagger$, forcing the mean to vanish.

For continuous-time Markov chains, the well‑known TUR reads
\begin{equation}
    \frac{\mathrm{Var}[\phi]}{\langle\phi\rangle^{2}}
    \;\ge\;
    \frac{2}{\langle\sigma\rangle}.
    \label{tur}
\end{equation}
This bound is sharp for the current-type observables.
In comparison, our bound is looser since $f$ in Eq.~\eqref{res} grows faster and thus
\(
    e^{-f(\langle\sigma\rangle)}
    \le
    {2}/{\langle\sigma\rangle}
  \)
for all $\langle\sigma\rangle>0$.
However, it was found that even in continuous-time Markov chains, Eq.~\eqref{tur} can be violated for observables other than current \cite{hasegawa2019ftur}.
By contrast, our TUR holds for arbitrary non-equilibrium dynamics as long as the DFT of Eq.~\eqref{core} is satisfied.
Indeed, whether they are stochastic (Markov jump or Langevin) or deterministic Hamiltonian evolution under cyclic driving, our derivation is independent of the underlying dynamics.

\subsection{Derivation}

We introduce a real-valued joint distribution \(P(\sigma,\phi)\) induced by
the path ensemble through the map
\mbox{
\(
\Gamma\mapsto(\sigma[\Gamma],\phi[\Gamma]),
\)
}
and call it a \emph{generalized FT distribution} if the underlying path ensemble satisfies the trajectory-level fluctuation
relation in Eq.~\eqref{core} and the observable satisfies the generalized
time-reversal condition in Eq.~\eqref{a}. Equivalently, the support of
\(P(\sigma,\phi)\) can be decomposed into time-reversal pairs
\begin{equation}
    \left(\sigma(\Gamma),\phi(\Gamma)\right)\longleftrightarrow\left(-\sigma(\Gamma),\phi(\Gamma^{\dagger})\right),
    \quad \phi(\Gamma)\phi(\Gamma^{\dagger})\le0,
    \label{TRpair}
\end{equation}
whose probability masses satisfy the following strong DFT:
\begin{equation}
    P\left(\sigma(\Gamma),\phi(\Gamma)\right)=e^{\sigma(\Gamma)} P\left(-\sigma(\Gamma),\phi(\Gamma^{\dagger})\right).
\end{equation}
Here, the time-reversed pair of observables, \(\phi(\Gamma)\) and \(\phi(\Gamma^\dagger)\), is constrained only through their signs. 
To handle such a distribution under weak constraints, our derivation employs a statistical bound for general real probability distributions.

Cantelli's inequality, also known as the one-sided Chebyshev inequality, provides an upper bound on the tail probability of a random variable~\cite{BoucheronLugosiMassart2013,Ghosh2002RelatedMarkov}.
Let $X$ be a real random variable with its expectation value $\mu=\langle X\rangle$ and variance $\sigma^2=\mathrm{Var}[X]$.  Then for any $t>0$,
\mbox{$\Pr\bigl(X - \mu \ge t\bigr)\;\le\;{\sigma^2}/(\sigma^2 + t^2).$}
Compared to the two‐sided Chebyshev inequality, Cantelli's inequality provides a strictly tighter estimate when only a single tail probability is of interest.
A mirrored bound can be obtained by changing the variable to $-X$,
\begin{equation}
\Pr(X-\mu\le -t)\le\frac{\sigma^2}{\sigma^2+t^2}.\label{eq:cantelli_lower_tail}
\end{equation}
We now apply Eq.~\eqref{eq:cantelli_lower_tail} to the observable
\(X=\phi(\Gamma)\). Denoting its mean and variance by \(\langle\phi\rangle\)
and \(\mathrm{Var}[\phi]\), respectively, we obtain
\begin{equation}
    \Pr\bigl(\phi-\langle\phi\rangle\leq -t\bigr)
    \leq
    \frac{\mathrm{Var}[\phi]}{\mathrm{Var}[\phi]+t^2}.
\end{equation}
Assuming \(\langle\phi\rangle>0\) and setting \(t=\langle\phi\rangle\), this
becomes
\begin{equation}
    \Pr(\phi\leq0)
    \leq
    \frac{\mathrm{Var}[\phi]}
    {\mathrm{Var}[\phi]+\langle\phi\rangle^2}.
\end{equation}
Rearranging yields the lower bound on the familiar inverse signal-to-noise ratio,
\begin{equation}
    \frac{\mathrm{Var}[\phi]}{\langle\phi\rangle^2}
    \geq
    \frac{\Pr(\phi\leq0)}{1-\Pr(\phi\leq0)}.
    \label{canteb}
\end{equation}
Note that since the sign convention of \(\phi\) is arbitrary, if \(\langle\phi\rangle<0\) then one may replace \(\phi\to-\phi\) without changing \(\mathrm{Var}[\phi]\) or \(\langle\phi\rangle^{2}\).
In other words, one can assume \(\langle\phi\rangle>0\) without loss of generality.

From here, we derive the TUR for generalized time-reversal observables [Eq.~\eqref{res}] by first finding the minimum possible tail \(\Pr(\phi\le0)\) in the generalized FT distribution class, and this minimum leads to the lower bound on the RHS of Eq.~\eqref{canteb}.
We proceed in three steps to find the minimum regarding the generalized FT distributions with fixed expectation values $\langle\sigma\rangle$ and $\langle\phi\rangle$:

\begin{enumerate}[label=(\arabic*)]
  \item
    We focus on distributions
    supported on exactly one time-reversal pair (two points in total) and 
    evaluate the minimum tail probability in this two-point class. 
    The minimum is given by the smaller probability of the two supports.

  \item 
    For a generalized FT distribution with finite support, we derive the lower bound on \(\Pr(\phi\le0)\).
    This is achieved by regarding the support as a set of time-reversal pairs studied in (1).
    After this, Jensen's inequality is applied. The lower bound for the finite case is obtained.
    It coincides with the minimum obtained for two-point distributions and cannot be further lowered.

  \item
    We show that, for any desired values \(\bar\sigma>0\) and \(\bar\phi>0\), there exists a two-point distribution that attains the obtained minimum. Hence, the minimum over all generalized distributions can always be realized by a two-point generalized distribution under the given conditions.

\end{enumerate}
After this, we derive a universal lower bound of this class by substituting the obtained minimum of $\Pr(\phi\leq0)$ into the SNR bound in Eq.~\eqref{canteb}.
It is expressed in the language of dissipative cost, namely entropy production.

\begin{figure}[htbp]
  \centering
  \def\svgwidth{0.75\columnwidth}
  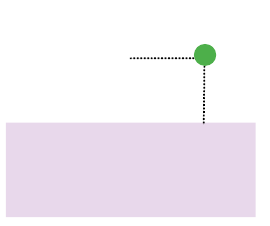
  \caption{
  Conceptual image of a generalized FT distribution.
  The support is composed of three time-reversal pairs.
  Each pair is indexed as $r=\{\Gamma_r^+,\Gamma_r^-\}$.
  The green pair is highlighted as an example.
  The probabilities within each pair are in the exponential ratio of Eq.~\eqref{core}, depicted by the sizes of the dots. 
  In a minimal configuration, the points with smaller probability have nonpositive $\phi$ and lie in the colored region.
  }
  \label{fig:support}
\end{figure}

Now, we consider a generalized FT distribution supported on exactly one
time-reversal pair and evaluate the minimum possible tail probability
\(\Pr(\phi\le 0)\).
We first consider a distribution whose support is
\begin{equation}
  (\sigma,\phi)=
  \begin{cases}
    (s,u),\\
    (-s,v),
  \end{cases}
  \qquad
  s>0,\quad uv\le 0 .
  \label{eq:single_pair_support}
\end{equation}
Here, \(s\) denotes the magnitude of the entropy production in the pair.
By the trajectory-level fluctuation relation in Eq.~\eqref{core}, the
normalized probability masses assigned to the two points are
\begin{equation}
  p(s)=\frac{e^s}{1+e^s},
  \qquad
  p(-s)=\frac{1}{1+e^s}.
  \label{eq:single_pair_masses}
\end{equation}
For \(s>0\), one has \(p(s)>p(-s)\).
This means that the negative-entropy point always carries the smaller probability. 
The minimum value of
\(\Pr(\phi\le 0)\) is obtained when only the negative point has
nonpositive \(\phi\) and equals
\begin{equation}
  p(-s)
  =
  \frac{1}{1+e^s}
  =: y(s).
  \label{yfrom}
\end{equation}

We next show that this two-point value gives a lower bound for any
generalized FT distribution.
Our approach is to decompose the support into time-reversal pairs.
Let \(P(\sigma,\phi)\) be a generalized FT distribution with finite
support, and fix
\(
  \langle \sigma\rangle=\bar\sigma>0,
  \,
  \langle \phi\rangle=\bar\phi>0 .
\)
We introduce \(\mathcal R\) as the set of all time-reversal
pairs with nonzero entropy production.
Then each element \(r\in\mathcal R\) is written as
\[
  r=\{\Gamma_r^+,\Gamma_r^-\}.
\]
Here, \(\Gamma_r^+\) and \(\Gamma_r^-\) denote the trajectories with
positive and negative entropy production in the pair \(r\):
\begin{equation}
  \Gamma_r^-=(\Gamma_r^+)^\dagger,
  \qquad
  -\sigma(\Gamma_r^-)=\sigma(\Gamma_r^+)>0.
\end{equation}
Thus, \(r\) is a pair of time-reversed trajectories with
sign-flipped entropy productions [Fig.~\ref{fig:support}].
For clarity, we assume that there are no vanishing trajectories with
\(\sigma=0\).
In Appendix~\ref{appendixc}, we show that this restriction does not affect the result.

For each pair \(r\), define its total probability mass by
\begin{equation}
  \mathcal P(r)
  :=
  \mathcal P(\Gamma_r^+)
  +
  \mathcal P(\Gamma_r^-).
  \label{eq:pair_mass}
\end{equation}
where \(\mathcal P(\Gamma_r^\pm)\) denotes the probability mass assigned to the
trajectory \(\Gamma_r^\pm\).
Conditioning on the pair \(r\), Eq.~\eqref{core} again gives the respective conditional masses
\begin{equation}
\left\{
\begin{aligned}
\Pr(\Gamma_r^+ \mid r)
&=
\frac{e^{\sigma(\Gamma_r^+)}}{1+e^{\sigma(\Gamma_r^+)}},\\
\Pr(\Gamma_r^- \mid r)
&=
\frac{1}{1+e^{\sigma(\Gamma_r^+)}} .
\end{aligned}
\right.
\label{conditional}
\end{equation}
Again, the conditional tail probability is locally minimized when only the point of negative entropy production contributes to this event. Hence,
\begin{align}
  \Pr(\phi\le 0 \mid r)
  &\ge
  \Pr(\Gamma_r^- \mid r)
  \nonumber\\
  &=
  \frac{1}{1+e^{\sigma(\Gamma_r^+)}}
  \nonumber\\
  &=
  y(\sigma(\Gamma_r^+)).
  \label{ydef}
\end{align}
Crucially, when the tail \(\Pr(\phi\leq0)\) is globally minimized, this condition should hold within every time-reversal pair.
Any deviation from such a configuration only increases the global tail probability, with each violating pair contributing the exact additional amount prescribed by Eq.~\eqref{core}.
Thus, using the law of total probability over the set of all pairs, we obtain
\begin{align}
  \Pr(\phi\le 0)
  &=
  \sum_{r\in\mathcal R}
  \Pr(\phi\le 0 \mid r)
  \mathcal P(r)
  \nonumber\\
  &\ge
  \sum_{r\in\mathcal R}
  y(\sigma(\Gamma_r^+))
  \mathcal P(r).
  \label{globallower}
\end{align}
Here, the summand is evaluated as a function of \(\sigma(\Gamma_r^+)\) since, for each pair \(r\), \(\Gamma_r^+\) denotes its member with positive entropy production.

We now impose the constraint on the mean entropy production.
Within a time-reversal pair with \(\sigma(\Gamma_r^+)=s>0\), the conditional mean entropy production is
\begin{align}
  s\,p(s)-s\,p(-s)
  &=
  s\frac{e^s-1}{e^s+1}
  \nonumber\\
  &=
  s\tanh\left(\frac{s}{2}\right)=:x(s).
\end{align}
Then the constraint \(\langle\sigma\rangle=\bar\sigma\) can be written as
\begin{equation}
  \sum_{r\in\mathcal R}
  x(\sigma(\Gamma_r^+))\,
  \mathcal P(r)
  =
  \bar\sigma .
  \label{subject_to}
\end{equation}
Then, the problem is equivalent to minimizing Eq.~\eqref{globallower} subject to Eq.~\eqref{subject_to}.
Since \(x(s)\) is strictly increasing for \(s>0\), its inverse \(x^{-1}\) exists, meaning that \(x\) is a one-to-one map onto \((0,\infty)\).
Furthermore, a direct calculation shows the composite function \(y\circ x^{-1}\) is convex on \((0,\infty)\) [see Appendix~\ref{appendixD}].
Let
\begin{equation}
  a_r
  :=
  x(\sigma(\Gamma_r^+)).
\end{equation}
Then using \(\sigma(\Gamma_r^+)=x^{-1}(a_r)\), Eq.~\eqref{globallower}
gives
\begin{align}
  \sum_{r\in\mathcal R}
  y(\sigma(\Gamma_r^+))
  \mathcal P(r)
  &=
  \sum_{r\in\mathcal R}
  (y\circ x^{-1})(a_r)
  \mathcal P(r)
  \nonumber\\
  &\ge
  (y\circ x^{-1})
  \left(
    \sum_{r\in\mathcal R}
    a_r
    \mathcal P(r)
  \right)
  \nonumber\\
  &=
  (y\circ x^{-1})
  \left(
    \sum_{r\in\mathcal R}
    x(\sigma(\Gamma_r^+))
    \mathcal P(r)
  \right)
  \nonumber\\
  &=
  y(x^{-1}(\bar\sigma)).
  \label{gousei}
\end{align}
Here, Jensen's inequality is applied in the second line; \mbox{$\mathbb{E}_{\mathcal R}[y\circ x^{-1}(a_r)]\geq y\circ x^{-1}(\mathbb{E}_{\mathcal R}[a_r])$.}
Although the support is finite, the function $y$ from Eq.~\eqref{yfrom} still bounds the tail probability.
Combining Eqs.~\eqref{globallower} and \eqref{gousei}, we obtain
\begin{equation}
  \Pr(\phi\le 0)
  \ge
  y(x^{-1}(\bar\sigma)).
  \label{eq:tail_bound_y}
\end{equation}
Although the support is finite, the function \(y\) from Eq.~\eqref{yfrom} still bounds the tail probability.

Finally, we show that the obtained lower bound is always achievable with a certain two-point generalized distribution.
Consider a generalized distribution \(P^\ast(\sigma,\phi)\) supported on one time-reversal pair:
\begin{equation}
  (\sigma,\phi)=
  \begin{cases}
    (s^\ast,u')\\
    (-s^\ast,-v')
  \end{cases}
  \qquad
  (u',v'\ge0),
\end{equation}
where \(s^\ast>0\) satisfies \(\bar\sigma=x(s^\ast)\).
Such \(s^\ast\) always exists for any given value \(\bar\sigma\) due to the monotonicity of \(x(s)\).
Then, \(P^\ast(\sigma,\phi)\) achieves the lower bound in Eq.~\eqref{gousei} with its tail probability:
\begin{align}
    p(-s^\ast)
    &=
    \frac{1}{1+e^{s^\ast}}
    \nonumber\\
    &=
    y(s^\ast)
    \nonumber\\
    &=
    y(x^{-1}(\bar\sigma)).
\end{align}
By construction, the two-point distribution \(P^\ast(\sigma,\phi)\) has the desired expectation value of \(\bar\sigma\).
Furthermore, it is always possible to choose \(u',v'\ge0\) so that \(P^\ast\) achieves the expectation value \(\bar\phi\) as well.
For example, taking
\begin{equation}
  u'=\bar\phi\,\frac{1+e^{s^\ast}}{e^{s^\ast}},
  \qquad
  v'=0
  \label{opt}
\end{equation}
gives
\begin{equation}
\langle\phi\rangle_{P^\ast}
  =
  p(s^\ast)u'-p(-s^\ast)v'
  =
  \bar\phi .
\end{equation}
Therefore, for any prescribed values
\(\bar\sigma>0\) and \(\bar\phi>0\), there exists a two-point generalized
FT distribution that achieves the lower bound on \(\Pr(\phi\le0)\).

From the preceding results, we obtain the following inequality:
\begin{align}
    \Pr(\phi\le0)
    &=
    \sum_{r\in\mathcal R}
    \Pr(\phi \le 0 \mid r)\mathcal P(r)
    \nonumber\\
    &\ge
    \sum_{r\in\mathcal R}
    y(\sigma(\Gamma_r^+))\mathcal P(r)
    \nonumber\\
    &\ge
    y(s^\ast)
    \nonumber\\
    &=
    \frac{1}{1+e^{s^\ast}}.
    \label{koko}
\end{align}

In the preceding argument, we saw that a two-point generalized distribution can always be constructed to achieve the lower bound on $\Pr(\phi\leq0)$, given fixed values of $\langle\sigma\rangle\,\text{and}\,\langle\phi\rangle$. 
Substituting this minimum possible tail into the SNR of Eq.~\eqref{canteb} yields the following form of TUR:
\begin{equation}
  \frac{\mathrm{Var}[\phi]}{\langle\phi\rangle^2}
  \;\ge\;
  \frac{y(s^\ast)}{1-y(s^\ast)}
  =
  \frac{1}{e^{s^\ast}},
\end{equation}
and in terms of the mean entropy production,
\begin{equation*}
  \frac{\mathrm{Var}[\phi]}{\langle\phi\rangle^2}
  \;\ge\;
    \frac{1}{e^{f(\langle\sigma\rangle)}}
    ,
    \quad
    f:=x^{-1}.
\end{equation*}
This is the main result in Eq.~\eqref{res}, and is valid for any generalized FT distribution.
In Appendix~\ref{appendixE}, we discuss the equality condition of this and see that there exists an optimized two-point distribution that saturates the bound.

\subsection{Extensions}

The novelty of our relaxed constraint is twofold.
First, the observables are no longer restricted to having symmetric magnitudes; \(|\phi(\Gamma^\dagger)|\ne|\phi(\Gamma)|\).
This allows us to describe situations where forward
and reverse trajectories may carry signals of different magnitudes.
An experimentally relevant example is a nonlinear readout mechanism. 
For instance, a sensing detector may respond with different sensitivities to positive and negative inputs~\cite{Momeni2025PhysicalNN,Williamson2020EOActivation,Li2023AllOpticalReLU,Xu2022PhotonicActivation}, leading to magnitude-asymmetric observed signals.
Second, the observable may discard information or vanish on a subset of trajectories, such as in the case of thresholding. 
In such cases, sub-threshold trajectories are mapped to
the same value, making the observable degenerate on those events. 
In particular, under the conventional constraint in Eq.~\eqref{typicallyimposed}, the observables can vanish only in the trivial case: $\phi(\Gamma)=\phi(\Gamma^\dagger)=0$.
By contrast, the relaxed constraint captures, for instance, observables designed to detect only a specific \emph{direction} of a quantity, such as work done \emph{on} the environment.
To illustrate this, consider the thresholded observable
\begin{equation}
    \phi(\Gamma)\;=\;\max\{\,W(\Gamma),\,0\}.
    \label{obsW}
\end{equation}
Under time reversal, the work changes its sign.
Hence, the opposite direction of the work is cut off in every pair, and we can see
\begin{equation}
    \phi(\Gamma)\,\phi(\Gamma^\dagger)
  =
  \max\{W,0\}\,\max\{-W,0\}
  =0,
  \label{liesinscope}
\end{equation}
which lies in the scope of our constraint $\phi(\Gamma)\,\phi(\Gamma^\dagger)\le0$.
Our TUR applies not only to this positive-work observable but to various kinds of properly defined observables with a \emph{direction} of interest.  
Later, we will demonstrate our TUR applied in this spirit.

Although Cantelli's inequality is sharp~\cite{LinBai2011ProbabilityInequalities} when the first two moments are specified, its extremal distribution, supported on two points, may lack physical relevance.
Here, let us remark that by imposing additional physicality on the distribution, we can derive an improved TUR bound.
Unimodality is such a physically mild assumption that still covers a broad class of observables.
We call a distribution unimodal if it has a single peak or mode $m$; more formally, when its cumulative distribution function is convex on $(-\infty, m]$ and concave on $[m, \infty)$.
It is a natural shape constraint in a wide-ranging stochastic modeling of practical interest~\cite{BalabdaouiJankowski2016,Inverse_Gaussian,CLTforAdditive} and covers many standard distribution families such as the Gaussian.

One-sided Vysochanskij--Petunin inequality~\cite{one-sidedVP,three_sigma,chevforuni} provides an upper bound on the tail probability of a unimodal real random variable. The additional unimodality assumption makes the bound tighter than Cantelli's inequality by a constant factor. Let $X$ be a real random variable with mean $\mu=\langle X\rangle$ and variance $\sigma^2=\mathrm{Var}[X]$. Then, for any $t>0$, $\Pr(X-\mu \le -t) \le 4\sigma^2 / [9(t^2+\sigma^2)]$ if $t^2 \ge 5\sigma^2/3$, and $\Pr(X-\mu \le -t) \le 4\sigma^2 / [3(t^2+\sigma^2)] - 1/3$ if $t^2 < 5\sigma^2/3$.
Rearranging these inequalities under the same argument as for the Cantelli bound, we obtain the following  SNR bound for a unimodal observable $\phi$,
\begin{equation}
\frac{\mathrm{Var}[\phi]}{\langle\phi\rangle^2}
\;\ge\;
\begin{cases}
\dfrac{9\,\Pr(\phi\le 0)}{\,4-9\,\Pr(\phi\le 0)\,}, & 0\le \Pr(\phi\le 0) \le \dfrac{1}{6},
\\[12pt]
\dfrac{1+3\,\Pr(\phi\le 0)}{\,3\bigl(1-\Pr(\phi\le 0)\bigr)\,}, & \dfrac{1}{6}\le \Pr(\phi\le 0) < 1.
\end{cases}
\end{equation}
Substituting the minimum distribution in Eq.~\eqref{koko} yields the following form of TUR:
\begin{equation}
\frac{\mathrm{Var}[\phi]}{\langle\phi\rangle^2}
\;\ge\;
\begin{cases}
\dfrac{4}{3\,e^{f(\langle\sigma\rangle)}}+\dfrac13,
& 0\le \langle\sigma\rangle \le \dfrac{2}{3}\ln 5,
\\[10pt]
\dfrac{9}{4\,e^{f(\langle\sigma\rangle)}-5},
& \langle\sigma\rangle \ge \dfrac{2}{3}\ln 5.
\end{cases}
\label{VPTUR}
\end{equation}
The inequalities have branches since the achievable minimum variance depends on the position of the mode relative to the tail threshold.
We see below that a certain class of non-unimodal variables also follows the improved bound of Eq.~\eqref{VPTUR}, in a way relevant to the context of generalized time reversal.

\section{Examples}

\subsection{Three-level quantum thermal machine}

\begin{figure}[tb]
  \centering
  \def\svgwidth{0.7\columnwidth}
  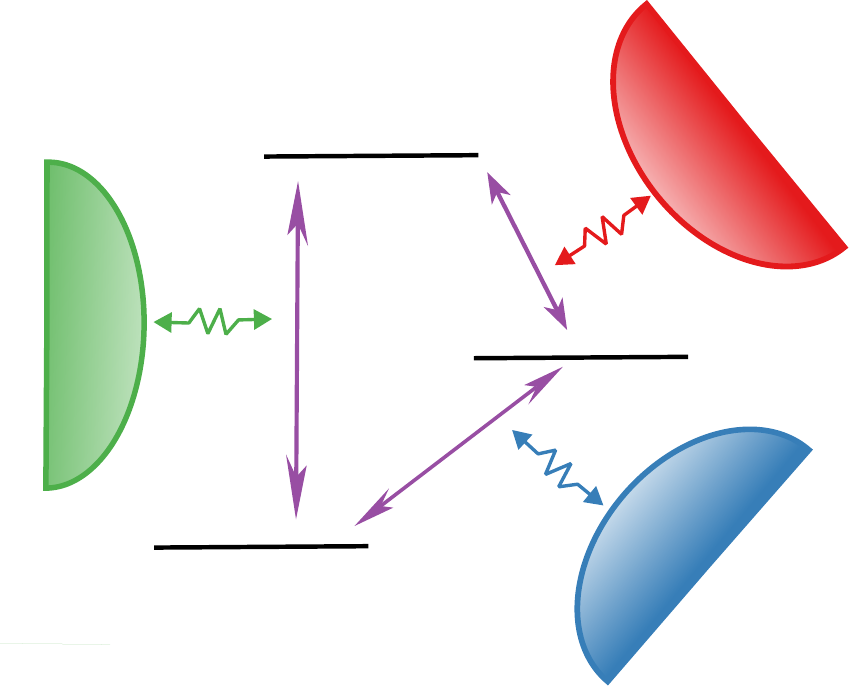
  \caption{Schematic diagram of a three-level quantum thermal machine. The machine is weakly coupled to each bath $\ell$, exchanging quantized energy $\hbar\omega_\ell$ in induced jumps (thick arrows). 
  In the refrigerator configuration, heat is extracted from the cold bath $\beta_1$ and dumped into the medium bath $\beta_3$ at the expense of energy supplied by the hot bath $\beta_2$.}
  \label{fig:3level0}
\end{figure}
We apply our TUR in Eq.~\eqref{res} to an autonomous quantum thermal machine [Fig.~\ref{fig:3level0}], particularly a three-level quantum refrigerator that operates solely with a hot bath as its energy source~\cite{Scovil1959,Alicki1979,GevaKosloff1994,BoukobzaTannor2006}.
Cooling plays a fundamental role in quantum technologies~\cite{tan2017_qcr,jones2020_ultralc,mitchison2019_qam}, making autonomous absorption machines a compelling class of systems for investigating thermodynamic precision.
A recent study in Ref.~\cite{HegdePottsLandi2025timeresolved} classified individual quantum jump trajectories of a quantum thermal machine into distinct thermodynamic cycles.
This allows for evaluating the machine's efficiency in performing its intended task, beyond simple average measures.

Following the presentation in Ref.~\cite{ManzanoPaule2018inopenquantum}, we represent the thermal machine's energy eigenstates as $\{\ket{g},\,\ket{e_A},\,\ket{e_B}\}$,
and Hamiltonian $\hat H_S = \hbar\omega_1\,\ket{e_A}\bra{e_A}
+\hbar(\omega_1+\omega_2)\,\ket{e_B}\bra{e_B},$
so that the frequency gaps are
\(\omega_{g\leftrightarrow A}=\omega_1\), \(\omega_{A\leftrightarrow B}=\omega_2\) and \(\omega_{g\leftrightarrow B}=\omega_3:=\omega_1+\omega_2\).
Each of these three transitions is weakly coupled to a bosonic reservoir \(\ell\in\{1,2,3\}\) at inverse temperature \(\beta_\ell=1/T_\ell\), inducing jumps described by ladder operators
$
\ket{g}\bra{e_A},\quad
\ket{e_A}\bra{e_B},\quad
\ket{g}\bra{e_B},
$
and jump rates $k^{(\ell)}_{\!\uparrow},k^{(\ell)}_{\!\downarrow}$ that satisfy the detailed balance condition:
\begin{equation}
    \frac{k^{(\ell)}_{\!\downarrow}}{k^{(\ell)}_{\!\uparrow}}
    \;=\;
    \frac{\gamma_\ell(n_\ell^{\rm th}+1)}{\gamma_\ell n_\ell^{\rm th}}
    \;=\;
    e^{\beta_\ell\hbar\omega_\ell},
\end{equation}
where
$
n_\ell^{\rm th}=1/(e^{\beta_\ell\hbar\omega_\ell}-1)
$ represents populations following a Bose-Einstein distribution, and $\gamma_\ell$ the spontaneous decay rate.
We consider a standard jump measurement under which the dynamics reduce to transitions between eigenstates, being formally equivalent to those of a classical Markov process~\cite{Hasegawa2020_QTUR_CM}.
Since we are interested in the non-equilibrium steady state \(\dot \pi=0\) the system reaches after sufficient time, the populations \(\pi_i\) of each eigen level here can be given by $\pi_g = [ e^{\theta_3}( 2 e^{\theta_1 + \theta_2} - 1 )-e^{\theta_1 + \theta_2}] / Z_\pi$, $\pi_{e_A} = [ e^{2 \theta_2} ( e^{\theta_1} - 2 ) + e^{\theta_3} ( 2 e^{\theta_2} - 1 ) ]/Z_\pi$, and $\pi_{e_B}=[e^{\theta_3} + e^{\theta_1 + \theta_2} - 2 ]/Z_\pi$, where $\theta_\ell:=\beta_\ell\hbar\omega_\ell$, and $Z_\pi = e^{2\theta_2}(-2 + e^{\theta_1})- 2 + e^{\theta_3}[ 2e^{\theta_2}(1+e^{\theta_1}) - 1].$
For clarity, we assumed the baths share an identical decay rate $\gamma$.
Then the heat currents at each bath can be denoted as
$
\dot Q_\ell=\hbar\omega_\ell\bigl(k^{(\ell)}_{\!\uparrow}\pi_{-} -k^{(\ell)}_{\!\downarrow}\pi_{+}\bigr),
$~\cite{Alicki1979,BoukobzaTannor2006,ManzanoPaule2018inopenquantum}
where \(\pi_{-}\, (\pi_{+})\) is the population of the lower (upper) energy level of the jump induced by the \(\ell\)th reservoir.
In this definition, $\dot Q_\ell$ is net positive when the machine absorbs it from the reservoir and negative when emitted.
A working refrigerator absorbs positive heat from cold and hot reservoirs, and emits it to the intermediate one. 
For a cyclic steady operation, the system entropy does not change on average.
Hence, the total entropy production is given by the entropy change of the baths and must be nonnegative.
This observation yields the condition that any autonomous refrigerator must satisfy to achieve its autonomy~\cite{ManzanoPaule2018inopenquantum}:
$
\Delta S_{\text{baths}} = -\beta_1 \hbar\omega_1 - \beta_2 \hbar\omega_2 + \beta_3 \hbar\omega_3 \geq 0.
$
In accordance with this, we can specify our parameters so that
\begin{equation}
    \omega_2 \;\ge\;
    \frac{\beta_1-\beta_3}{\beta_3-\beta_2}\,\omega_1.
    \label{cond}
\end{equation}
When this inequality is true, the average heat flux entering from the first reservoir remains positive.
Since this system is in a non-equilibrium steady state (NESS) and there is no external driving force, it admits the trajectory-level statistical relation in Eq.~\eqref{core}.

Next, we define our observable with $Q_1$ in the same manner as in Eq.~\eqref{obsW}:
\begin{equation}
    \phi(\Gamma)\;=\;\max\{\,Q_1(\Gamma),\,0\}.
    \label{obsQ}
\end{equation}
where $Q_1(\Gamma)$ denotes the total heat transferred in the transitions between \mbox{$\ket{g}\leftrightarrow \ket{e_A}$} along a trajectory $\Gamma$.
The quantity $\phi(\Gamma)$ has a magnitude only when the net absorption $Q_1(\Gamma)$ is positive.
Hence, Eq.~\eqref{obsQ} depicts the extraction of heat from the cold reservoir, which is the primary function of the cooling machine considered here. 
In this setting, our TUR is applicable to the observable of generalized time-reversal $\phi(\Gamma)$ that satisfies the analog of Eq.~\eqref{liesinscope}.
Note that by considering both \emph{directions} of heat exchanged with the cold reservoir, this observable becomes an antisymmetric one $Q_1=:\phi'$ whose time-reversal is
$\phi'(\Gamma^\dagger) = -Q_1=-\phi'(\Gamma)$.
\begin{figure}[tb]
    \centering
    \includegraphics[width=1.0\linewidth]{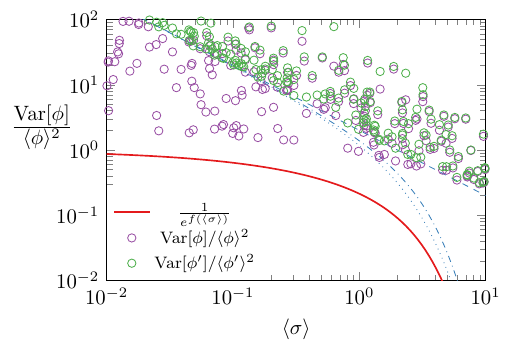}
    \caption
    {Results of computer simulation for the thermal machine.
    $\mathrm{Var}[\varphi]/\langle\varphi\rangle^2$ as functions of $\langle\sigma\rangle$, where $\gamma\in[0.01, 0.1],\beta_1\in[1.0,10.0],\,\beta_2\in[0.01, 0.1],\,\beta_3\in[0.1,1.0], \omega_1\in[1.0, 2.0], \omega_2\in[\kappa\omega_1, \kappa\omega_1+45]$ for $\kappa = (\beta_1-\beta_3)/(\beta_3-\beta_2)$ from Eq.~\eqref{cond}.
    The purple circles show the relative variance of $\phi$ obeying generalized time-reversal.
    The green ones show that of $\phi'$ with standard time-antisymmetry.
    The blue dashed lines denote the lower bounds $2/\langle\sigma\rangle$ and $2/({e^{\langle\sigma\rangle}-1})$ [Eqs.~\eqref{tur} and \eqref{FTUR}], and the dotted line denotes $h(\langle\sigma\rangle)$ [Eq.~\eqref{preEFTUR}].
    }
    \label{fig:resplot}
\end{figure}

We randomly select $\gamma, \omega_1, \omega_2, \beta_1, \beta_2, \beta_3$ in a way that the cooling condition of Eq.~\eqref{cond} is met [see the caption in Fig.~\ref{fig:resplot}].
The configuration also establishes the basis for the correctness of the model, such as weak coupling approximation $\gamma\ll\omega$, which ensures stable energy states.
We calculate $\langle\sigma\rangle,\,\text{and} \,\mathrm{Var}[\varphi]/\langle\varphi\rangle^2$ ($\varphi=\phi\;\text{or}\;\phi'$) along $50000$ trajectories of length $1000.0$ for many times and plot $\mathrm{Var}[\varphi]/\langle\varphi\rangle^2$ as a function of the observed entropy production $\langle\sigma\rangle$ in Fig.~\ref{fig:resplot}.

We provide a comparison of our bound with other TURs presented in the Appendix.
We can see our TUR bound is valid for both observables $\phi$ and $\phi'$, but the original TUR in Eq.~\eqref{tur} and FTUR in Eq.~\eqref{FTUR} are violated by the generalized observable $\phi$. 
This is because the current-type TUR and the FTUR are consequences of the strict time-antisymmetry of the considered observables.
The EFT bound of Eq.~\eqref{EFTUR} is depicted by the green dashed line.
Note that for $\phi'$, EFT holds in our setting;
$P(\sigma,\phi')/P(-\sigma,-\phi')=e^\sigma$, as the interaction is sufficiently weak and transient
\cite{timpanaro2019tur_exchange,hasegawa2019ftur} (see also \cite{jarzynski2004heat_exchange,SaitoUtsumi2008PRB,LandiKarevski2016,Lahiri2015notSoCommon}).
Still, the EFT bound loses its validity on $\phi$ for the same reason outlined above.

\subsection{One-dimensional diffusion channel}

\begin{figure}[tb]
  \centering
  \def\svgwidth{0.85\columnwidth}
  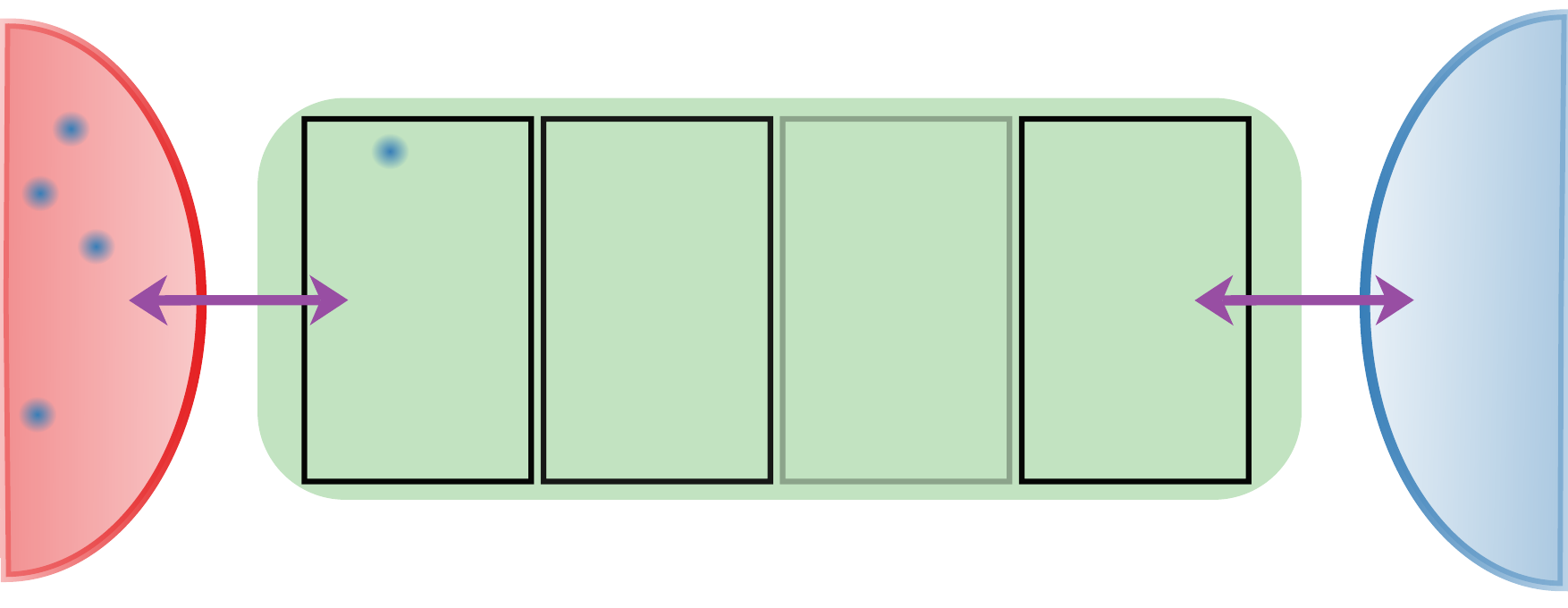
\caption{
Schematic diagram of the one-dimensional diffusion channel coupled to two
particle reservoirs. 
Through the channel, the local density imbalance of \(x_i\) acts as the driving force for particle transport.
The arrows denote the bond currents \(dQ_j\) flowing 
between adjacent cells, which is subject to Gaussian white thermal noise.
}
  \label{fig:oneDimChnl}
\end{figure}

Our second result in Eq.~\eqref{VPTUR} assumes unimodality of the underlying observable. 
Based on this assumption, we investigate our TURs on an autonomous transport model~\cite{Spohn1991,Bertini2015MFT}: a one-dimensional diffusion channel coupled to two particle reservoirs [Fig.~\ref{fig:oneDimChnl}]. 

We discretize the channel into \(N\) coarse-grained cells and denote the
density state by
\(X_t := (x_1(t),\dots,x_N(t))^\top \in \mathbb{R}^N\). The left and right
reservoirs impose effective boundary concentrations \(\mu_L\) and \(\mu_R\),
respectively. When \(\mu_L\neq\mu_R\), this concentration imbalance drives a
stationary current through the system~\cite{BodineauDerrida2004,Bertini2005Current}.
The dynamics are described by bond currents \(dQ_j(t)\),
\(j=0,1,\dots,N\), each consisting of a deterministic Fick-type contribution
and intrinsic thermal white noise~\cite{LandauLifshitzFluid,Bell2007LLNS,Bertini2015MFT}. In the bulk,
\(dQ_j(t)=\kappa(x_j(t)-x_{j+1}(t))\,dt
+\sqrt{2\kappa/\beta}\,dW_j(t)\) for \(j=1,\dots,N-1\). At the boundaries,
\(dQ_0(t)=k_L(\mu_L-x_1(t))\,dt+\sqrt{2k_L/\beta}\,dW_0(t)\) and
\(dQ_N(t)=k_R(x_N(t)-\mu_R)\,dt+\sqrt{2k_R/\beta}\,dW_N(t)\). Here
\(k_L,k_R>0\) are the reservoir--channel couplings, \(\kappa>0\) is the bulk
diffusivity, \(\beta\) is the inverse temperature, and
\(\{W_j\}_{j=0}^N\) are independent Wiener processes, characterized by independent Gaussian increments.
Mass conservation gives \(dx_i=dQ_{i-1}-dQ_i\) for \(i=1,\dots,N\). This
defines a linear Langevin system of the Ornstein--Uhlenbeck type,
\(dX_t=(AX_t+b)\,dt+B\,dW_t\), with time-independent coefficients. Thus the
dynamics are autonomous, and for \(\mu_L\neq\mu_R\) the system is in NESS. We initialize the system in this steady state.
The medium entropy production along a trajectory \(\Gamma\) is written in the
Stratonovich form as
\begin{equation}
\Sigma_{\mathrm{med}}[\Gamma]
:=
\beta \sum_{j=0}^{N} \int_0^T F_j(X_t)\circ dQ_j(t),
\label{eq:Sigma_med_force_flux}
\end{equation}
where the conjugate affinities are
\(F_0(X)=\mu_L-x_1\), \(F_j(X)=x_j-x_{j+1}\) for \(j=1,\dots,N-1\), and
\(F_N(X)=x_N-\mu_R\). This is the diffusion-limit expression of local detailed
balance.

Lastly, we consider an antisymmetric observable whose distribution is unimodal:
\begin{equation}
A_T:=\frac{1}{N-1}\sum_{j=1}^{N-1}\int_0^T dQ_j(t),
\label{eq:rectified}
\end{equation}
namely, bond-averaged integrated current. 
Since the dynamics is linear and each \(dQ_j(t)\) is an affine functional, \(A_T\) is Gaussian for every finite \(T>0\).
From this, We define a generalized time reversal observable going through a nonlinear readout mechanism;
\begin{equation}
    \phi(\Gamma)=
    \begin{cases}
        \alpha_-\, A_T & \text{if } A_T \leq 0\\
        \alpha_+\, A_T & \text{if } A_T > 0\\
    \end{cases}
    \quad \alpha_-, \,\alpha_+ >0.
    \label{leakyOBS}
\end{equation}
This piecewise asymmetric readout is natural in situations where the measured signal is processed through channels with different effective gains, for instance, as can occur in hardware neural systems with activation stages (see, e.g., recent work on physical neural networks~\cite{Momeni2025PhysicalNN}, memristive neuromorphic hardware~\cite{Aguirre2024MemristiveANN}, and photonic neural activations~\cite{Li2023AllOpticalReLU,Williamson2020EOActivation}).
Mathematically, it can be viewed as an extension of the parametric rectified linear unit or PReLU~\cite{He2015PReLU}.
This asymmetric gain breaks the time antisymmetry of the observable $\phi$, such that \mbox{$\phi(\Gamma)\phi(\Gamma^\dagger)=-\alpha_+\alpha_-A_T^2\le 0$}.
We point out that for the case of \(\alpha_-<\alpha_+\), the distribution is no longer unimodal.
Nevertheless, for the class studied here, Eq.~\eqref{VPTUR} is still applicable; see Appendix~\ref{app:vp_bound}.

\begin{figure}
    \centering
    \includegraphics[width=1.0\linewidth]{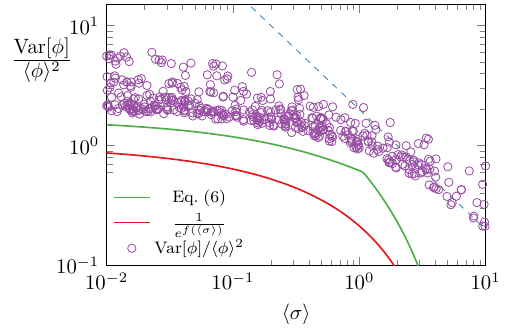}
    \caption
    {Results of computer simulation for the diffusion channel. $\mathrm{Var}[\phi]/\langle\phi\rangle^2$ as a function of $\langle\sigma\rangle$.
    Parameters are swept over \(\beta\in[10^{-0.3},10^{1}]\), \(k_L,\kappa,k_R\in[10^{-1},10^{1}]\), and \((\mu_L-\mu_R)\in[10^{-0.5},10^{1}]\), while the mean offset \((\mu_L+\mu_R)/2\) is drawn uniformly from a bounded interval.
    The purple circles show the relative variance of \(\phi\), with five sets of parameters $(\alpha_-, \alpha_+) \in [
        (0.01, 1.5),
        (0.03, 2.0),
        (0.1, 1.0),
        (2.0, 10.0),
        (5.0, 15.0)]$.
    The blue dashed lines denote the well-known lower bound $2/\langle\sigma\rangle$.
    }
    \label{fig:resplot2}
\end{figure}

We test the bound of Eq.~\eqref{VPTUR} for the observable \(\phi\) in Eq.~\eqref{leakyOBS} with \(N=6\) and the parameters $(\alpha_{-}, \alpha_+)$ in the caption of Fig.~\ref{fig:oneDimChnl}. The system is initialized in the steady state, and the augmented process \((X_t,A_t)\) is simulated by exact discretization. For each realization, we randomly generate the parameters \((\beta,k_L,\kappa,k_R,\mu_L,\mu_R)\) and estimate \(\mathrm{Var}[\phi]/\langle\phi\rangle^2\), which are plotted as a function of \(\langle\sigma\rangle\) in Fig.~\ref{fig:resplot2}. We generate \(4000\) trajectories using \(60\) exact time steps over a wide range of $T$. 

We compare the conventional TUR \(2/\langle\Sigma\rangle\), our main result in Eq.~\eqref{res}, and our second result in Eq.~\eqref{VPTUR}. As shown in Fig.~\ref{fig:resplot2}, the relative variance of the observable \(\phi\) is more tightly bounded by the Vysochanskij–Petunin bound, whereas the Cantelli bound is looser in this case since it only utilizes the finiteness of the first two moments. 
However, the second bound in Eq.~\eqref{VPTUR} is not saturated.
This is expected since the two-point extremal distribution of Eq.~\eqref{koko} falls outside the unimodal class. 
Thus we present our result as a sufficient but not a saturable lower bound for the family of observables considered here.

\section{Conclusion}
In this paper, we have derived two lower bounds on the uncertainty of observables that exhibit the generalized time-reversal symmetry.
We did so by delivering pure statistical arguments on the properties of the generalized FT distribution.
The resulting bounds were applied to capture the thermodynamic constraints on the operations of a quantum refrigerator and a diffusion channel.

Notably, our framework covers a broader class of nontrivial fluctuating quantities. It thereby provides a new direction of generalization, encompassing observables with greater flexibility and practical relevance.
This includes signals subject to partial information loss due to rectifications.
Furthermore, we showed that the resulting bound can be improved whenever additional structural information about the distribution is available.
We demonstrated it by rewriting the SNR bound with an alternative concentration inequality.

The significance of the present work is also conceptual. It demonstrates that constraints imposed by irreversible time evolution remain tractable even for generalized observables. By characterizing these observables through their distributional freedom, our approach successfully captures their broader flexibility. In this sense, it relies on a shift in the formulation of TURs from the classical pathwise perspective to a coarse-grained, distribution-level one. The present work shows that this level of granularity is effective for evaluating observables with the condition of generalized time-reversal.

\appendix
\section{FLUCTUATION THEOREM UNCERTAINTY RELATION}
Reference~\cite{hasegawa2019ftur} derived an exponential TUR directly from fluctuation theorem symmetries.
The derivation depends only on a strong DFT for strictly time-antisymmetric observables, and the obtained bound takes the form of
\begin{equation}
    \frac{\mathrm{Var}[\phi]}{\langle \phi\rangle^2}\ge\,\frac{2}{e^{\langle\sigma\rangle}-1}.
    \label{FTUR}
\end{equation}
Any system that has a joint symmetry of the form:
$P(\sigma,\phi)=e^\sigma P(-\sigma,-\phi)$
admits this fluctuation theorem uncertainty relation (FTUR), including both continuous and discrete Markov processes under a periodic protocol \cite{hasegawa2019ftur}.

\section{THERMODYNAMIC UNCERTAINTY RELATION IN EFT SCENARIO}
 
In Ref.~\cite{timpanaro2019tur_exchange}, the authors considered charge (integrated currents) exchange between multiple reservoirs that admits the exchange fluctuation theorem (EFT)~\cite{jarzynski2004heat_exchange,Andrieux2009NJP,EspositoHarbolaMukamel2009RMP}.
In this setting, the observable of interest \(\phi\) was defined as a linear combination
of charges \(Q_i\), and the entropy production takes the form of
\(\sigma = \sum_i A_i Q_i\),
where \(A_i\) are the corresponding thermodynamic affinities \cite{timpanaro2019tur_exchange}.
They then introduced the joint distribution \(P(\sigma,\phi)\) for these time-antisymmetric quantities.
From all the probability distributions constrained by EFT, they identified the one where \(\mathrm{Var}[\phi]\) is minimized and thus the lower bound for the observable's uncertainty is given.
Their result shows that, for fixed means of the observable and entropy production $\langle \sigma\rangle$ and $\langle \phi\rangle$, the two-point distribution, namely $(\sigma,\phi)=\{(a,b), (-a,-b)\}$, has the smallest variance of $\phi$.
Parameters $a$ and $b$ are chosen to enforce both the EFT symmetry and the given first moments; 
$\langle \sigma\rangle = a\tanh(a/2)$ and $\langle \phi\rangle = b\tanh(a/2)$.
Then, $P_{\rm min}$ has the smallest variance satisfying \begin{equation}
    \;\mathrm{Var}[\phi]=\langle \phi\rangle^2\,h(\langle \sigma\rangle) \;\text{with}\;
    h(x)=\mathrm{csch}^2[g(x/2)],
    \label{preEFTUR}
\end{equation}
where $\mathrm{csch}(x)$ is the hyperbolic cosecant and $g$ is the inverse of $x\tanh(x)$.
Accordingly, the TUR in EFT scenarios reads:
\begin{equation}
    \frac{\mathrm{Var}[\phi]}{\langle \phi\rangle^2}\ge\,h(\langle \sigma\rangle).
    \label{EFTUR}
\end{equation}

\section{VANISHING TRAJECTORY}
\label{appendixc}

In the derivation of the minimum achievable tail bound
[Eq.~\eqref{koko}], we assumed that the support contains no
vanishing trajectories with \(\sigma=0\). We show here that this
restriction does not affect the minimization of the tail probability.

Let the support be decomposed into nonvanishing time-reversal pairs
\(r=\{\Gamma_r^+,\Gamma_r^-\}\), with \(\sigma(\Gamma_r^+)>0\), together with
a possible vanishing sector. For each nonvanishing pair, we write
\[
  \mathcal P(r)
  =
  \mathcal P(\Gamma_r^+)+\mathcal P(\Gamma_r^-).
\]

For a vanishing trajectory $\Gamma_0$, the time-reversed
points have equal probability. Hence, under the generalized
time-reversal constraint, the smallest possible conditional contribution
to the event \(\{\phi\le0\}\) is \(1/2\). Therefore, including the
vanishing sector, the lower bound corresponding to Eq.~\eqref{globallower}
becomes
\begin{align}
  \Pr(\phi\le0)
  &\ge
  \sum_{r\in\mathcal R}
  y(\sigma(\Gamma_r^+))\mathcal P(r)
  +
  \frac{1}{2}
  \sum_{\Gamma_0}
  \mathcal P(\Gamma_0)
  \nonumber\\
  &=
  \sum_{r\in\mathcal R}
  y(\sigma(\Gamma_r^+))\mathcal P(r)
  +
  \frac{1}{2}
  \left(
    1-
    \sum_{r\in\mathcal R}
    \mathcal P(r)
  \right).
  \label{Cappendix}
\end{align}
The summation is taken over the set of all nonvanishing time-reversal
pairs \(\mathcal R\).

It remains to check that the right-hand side cannot be made smaller by
assigning a nonzero probability mass to the vanishing sector. Define
\begin{equation}
  m
  :=
  \sum_{r\in\mathcal R}
  \mathcal P(r),
  \qquad
  0\le m\le 1 .
\end{equation}
The total mass of the vanishing sector is then \(1-m\). Since the
vanishing sector contributes no entropy production, the constraint on the
mean entropy production reads
\begin{equation}
  \sum_{r\in\mathcal R}
  x(\sigma(\Gamma_r^+))
  \mathcal P(r)
  =
  \bar\sigma ,
  \label{eq:vanishing_mean_constraint}
\end{equation}
where
\(
  x(s)=s\tanh\left({s}/{2}\right).
\)
For \(m>0\), applying Jensen's inequality to the normalized weights
\(\mathcal P(r)/m\) gives
\begin{align}
  \sum_{r\in\mathcal R}
  y(\sigma(\Gamma_r^+))
  \mathcal P(r)
  &\ge
  m\,
  (y\circ x^{-1})
  \left(
    \frac{\bar\sigma}{m}
  \right).
\end{align}
Thus, for fixed \(\bar\sigma>0\), the lower bound in
Eq.~\eqref{Cappendix} is bounded from below by
\begin{equation}
  F(m)
  :=
  m\,
  (y\circ x^{-1})
  \left(
    \frac{\bar\sigma}{m}
  \right)
  +
  \frac{1-m}{2}.
  \label{eq:Fm_vanishing}
\end{equation}
We now show that \(F(m)\) is minimized at \(m=1\). Let
\(g:=y\circ x^{-1}\) and set \(z=\bar\sigma/m\). Then the derivative is
\(
  F'(m)
  =
  g(z)-z g'(z)-{1}/{2}.
\)
Writing \(z=x(s)\), a direct calculation gives
\begin{equation}
  F'(m)
  =
  -
  \frac{
    \tanh^2(s/2)
  }{
    2\tanh(s/2)+s\,\mathrm{sech}^2(s/2)
  }
  \le 0 .
\end{equation}
Hence \(F(m)\) is nonincreasing in \(m\), and its minimum is attained at
the largest possible value \(m=1\). This is precisely the case in which
the vanishing sector has zero total probability:
\begin{equation}
  \sum_{\Gamma_0}\mathcal P(\Gamma_0)=0 .
\end{equation}
Equivalently,
\begin{equation}
  \sum_{r\in\mathcal R}
  \left[
    \mathcal P(\Gamma_r^+)
    +
    \mathcal P(\Gamma_r^-)
  \right]
  =
  1 .
  \label{novanishing}
\end{equation}
Substituting Eq.~\eqref{novanishing} into Eq.~\eqref{Cappendix}
recovers the lower bound in Eq.~\eqref{globallower}. Therefore,
introducing vanishing trajectories cannot decrease the minimum attainable
value of \(\Pr(\phi\le0)\). This justifies the simplification made in the
main text: in finding the distribution that minimizes the tail
probability, it suffices to consider distributions with no vanishing
trajectories.

An additional comment is on the equilibrium case of zero average entropy production: $\langle\sigma\rangle=0$.
In the main text, we excluded it by setting \mbox{$\langle\sigma\rangle=\bar\sigma>0$}.
We did so since zero entropy production is realized by the trivial distribution supported only on vanishing trajectories.

\section{CONVEXITY OF THE COMPOSITE FUNCTION}\label{appendixD}

In deriving Eq.~\eqref{koko}, we used the convexity of the composite function \mbox{$y\circ x^{-1}(a)$}.
The proof of this property will be provided by the direct calculation of its second derivative.

For $s>0$, let
$$y(s):=\frac{1}{1+e^{s}},\qquad x(s):=s\tanh\left(\frac{s}{2}\right).$$
Then, 
\begin{align}
  y'(s)&=-\frac{e^{s}}{(1+e^{s})^{2}}\nonumber \\
y''(s)&=\frac{e^{s}\!\left(e^{2s}-1\right)}{(1+e^{s})^{4}}\quad(\ge0),  
\end{align}
and
\begin{align}
  x'(s)&=\tanh\left(\frac{s}{2}\right)+\frac{s}{2}\frac{\cosh^{2}(\frac{s}{2})-\sinh^{2}(\frac{s}{2})}{\cosh^{2}(\frac{s}{2})} \nonumber \\
  &=\tanh\left(\frac{s}{2}\right)+\frac{s}{2}\operatorname{sech}^{2}\left(\frac{s}{2}\right)\quad(>0) \nonumber \\ 
  x''(s)&=\operatorname{sech}^{2}\left(\frac{s}{2}\right)\Bigl(1-\frac{s}{2}\tanh\left(\frac{s}{2}\right)\Bigr).
\end{align}
The composite function is $f(a)=y\circ x^{-1}(a)$ and we write $s=x^{-1}(a)$.
By the chain rule,
\begin{align}
    f'(a)&=y'(s)\,\frac{ds}{da}=\frac{y'(s)}{x'(s)}, \nonumber \\
    f''(a)&=\frac{y''(s)\,x'(s)-y'(s)\,x''(s)}{\bigl(x'(s)\bigr)^{3}}.
\end{align}
Hence,
\begin{align}
f''(a)
&=C
\Biggl[
\left(e^{2s}-1\right)\!\left(\tanh\left(\frac{s}{2}\right)
+\frac{s}{2}\operatorname{sech}^{2}\left(\frac{s}{2}\right)\right) \nonumber \\
&\hspace{0.5cm} +(1+e^{s})^{2}\operatorname{sech}^{2}\left(\frac{s}{2}\right)
\left(1-\frac{s}{2}\tanh\left(\frac{s}{2}\right)\right)
\Biggr],  
\end{align}
where $C={e^{s}}/{(1+e^{s})^{4}\bigl(x'(s)\bigr)^{3}} >0$.
The first term in the bracket is positive, but the second term is not because there exists a constant $s_0$ such that for $s>s_0$, the factor $\left(1-\frac{s}{2}\tanh\left(\frac{s}{2}\right)\right)$ becomes negative. 
However, the positive term, proportional to $e^{2s}-1$, dominates there.
Hence, the bracket remains positive and $f''(a)>0$ for all $a>0$.
Since the second derivative is positive everywhere, the composite function $y\circ x^{-1}(a)$ is strictly convex on the domain.

\section{EQUALITY CONDITION OF CANTELLI'S INEQUALITY}\label{appendixE}
We employed Cantelli's inequality in the derivation of our TUR; here we investigate the case where Eq.~\eqref{canteb} becomes equality.
The inequality originates from the chain for arbitrary \(t>0\) and \(a\ge0\):
\begin{align}
    \Pr\bigl(X-\mu\ge t\bigr)
    &\le
    \Pr\bigl((X-\mu+a)^2\ge (t+a)^2\bigr)\nonumber\\
    &\le
    \frac{\mathbb{E}[(X-\mu+a)^2]}{(t+a)^2}.
    \label{chain}    
\end{align}
The first relation is set inclusion and the second is Markov's inequality applied to the nonnegative random variable \((X-\mu+a)^2\).
To have equality throughout Eq.~\eqref{chain}, each equality must hold.

Since we have the set relation
\begin{align}
    \{(X-\mu+a)^2\ge (t+a)^2\}\hspace{45mm}\nonumber\\=\{X-\mu\ge t\}\cup\{X-\mu\le-(t+2a)\},\nonumber
\end{align}
the inclusion equality holds when
\begin{equation}    
    \Pr\bigl(X-\mu\le-(t+2a)\bigr)=0,
    \label{inclusioneq}
\end{equation}
Hence, the two probabilities equate when there is no mass below \(\mu-(t+2a)\).

For a nonnegative \(Z\) and \(c>0\), Markov's inequality \( \Pr(Z\ge c)\le\mathbb{E}[Z]/c\) is an equality if and only if \(Z\) is supported on \(\{0,c\}\) \cite{LinBai2011ProbabilityInequalities}. 
Applied with \(Z=(X-\mu+a)^2\) and \(c=(t+a)^2\), equality holds when
\begin{equation}
    X-\mu+a\in\{0,\pm(t+a)\}.
    \label{markoveq}
\end{equation}
Combining Eq.~\eqref{inclusioneq} and Eq.~\eqref{markoveq}, \(X-\mu+a\le-(t+a)\) is required to have zero mass, so
the only possible values of \(X-\mu\) are
\begin{equation}
    X-\mu\in\{-a,\;t\}.
    \label{moto}
\end{equation}
Changing the random variable $X$ to $-X$, Eq.~\eqref{moto} becomes
\begin{equation}
    -X+\mu\in\{-a,\;t\}\quad\text{i.e.,}\quad X\in\{\mu+a,\;\mu-t\}.\nonumber
\end{equation}
Since the mean is fixed as $\mathbb{E}[X]=\mu$, we can express the following with variance:
\begin{equation}
\Pr(X=\mu-t)=\frac{\sigma^2}{\sigma^2+t^2},\quad a=\frac{\sigma^2}{t}.
\end{equation}

Therefore, in our case of $t=\mu$, the equality is achieved by a two-point distribution 
\begin{equation}
  \phi=
  \begin{cases}
    0 &\text{with probability} \quad \dfrac{\sigma^2}{\sigma^2+\mu^2} \\
    \mu+\frac{\sigma^2}{\mu} &\text{with probability} \quad \dfrac{\mu^2}{\sigma^2+\mu^2}.
  \end{cases}
  \label{cantell2}  
\end{equation}
The question is whether there exists a generalized FT distribution in this configuration.
It turns out that we can make such a distribution from the two-point distribution we constructed in the main text.
In the minimization of $\Pr(\phi\leq0)$, the tail probability was shown to be achieved by a two-point distribution 
\begin{equation}
  (\sigma,\phi)=
  \begin{cases}
    (-s^\ast,-v')\\
    (s^\ast,u')
  \end{cases}
  (u',v'\ge0)
  .
\end{equation}
They hold the lower bound for the relative variance;
\begin{equation}
  \frac{\mathrm{Var}[\phi]}{\langle\phi\rangle^2}
  \;\ge\;
    \frac{1}{e^{s^\ast}}.\nonumber
\end{equation}
We optimize this by setting $v'=0$ and $u'e^{s^\ast}/(1+e^{s^\ast})=\bar\phi$.
This is the choice made in Eq.~\eqref{opt}.
In this configuration, the distribution minimizes the variance while maintaining the mean.
After optimization, it reads
\begin{equation}
  (\sigma,\phi)=
  \begin{cases}
    (-s^\ast,\;0)&\text{with probability} \quad \dfrac{1}{(1+e^{s^\ast})}\\
    (s^\ast,\;\dfrac{(1+e^{s^\ast})\bar\phi}{e^{s^\ast}}\;)&\text{with probability} \quad \dfrac{e^{s^\ast}}{(1+e^{s^\ast})}.
  \end{cases}
  \vspace{4pt}
\end{equation}
Now, it can be shown that this distribution also satisfies the configuration of Eq.~\eqref{cantell2}.
First, one can check the variance of this distribution: $\sigma^2=\bar\phi^2/e^{s^\ast}$.
Then it is observed that
\begin{align}
    \dfrac{(1+e^{s^\ast})\bar\phi}{e^{s^\ast}}=\bar\phi+\frac{\bar\phi}{e^{s^\ast}}=\mu+\frac{\sigma^2}{\mu},\nonumber
\end{align}
and accordingly,
\begin{align}
    \dfrac{1}{(1+e^{s^\ast})}=\dfrac{\dfrac{\bar\phi^2}{e^{s^\ast}}}{\left(\bar\phi^2+\dfrac{\bar\phi^2}{e^{s^\ast}}\right)}=\dfrac{\sigma^2}{\sigma^2+\mu^2},\nonumber\\
    \dfrac{e^{s^\ast}}{(1+e^{s^\ast})}=\dfrac{{\bar\phi^2}}{\left(\bar\phi^2+\dfrac{\bar\phi^2}{e^{s^\ast}}\right)}=\dfrac{\mu^2}{\sigma^2+\mu^2}.\nonumber
\end{align}
Thus, the optimized distribution has the form of Eq.~\eqref{cantell2} and meets the equality condition.
Therefore, it saturates the lower bound for the relative variance derived in the main text;
\begin{equation}
  \frac{\mathrm{Var}[\phi]}{\langle\phi\rangle^2}
  = \frac{1}{e^{s^\ast}}.    
\end{equation}
We saw that a distribution can be constructed within the class of generalized FT distributions to attain the equality of Cantelli's inequality, and such a distribution also gives the lower bound for the scaled variance.
This leads us to conclude that the TUR of our main result [Eq.~\eqref{res}] is saturable.

\section{VYSOCHANSKIJ--PETUNIN TYPE BOUND FOR A NON-UNIMODAL CLASS}
\label{app:vp_bound}
In this appendix, we show that the Vysochanskij--Petunin-type bound in Eq.~\eqref{VPTUR} holds for a certain class of non-unimodal families. For the purpose of the present paper, it is sufficient to prove the result for the Gaussian-based piecewise-linear observable at
the threshold zero.

Without loss of generality, let
\begin{equation}
Y\sim \mathcal N(\mu,\sigma^2),\qquad \mu>0,
\end{equation}
and define
\begin{equation}
X=
\begin{cases}
\alpha_-Y, & Y\le 0,\\[1mm]
\alpha_+Y, & Y>0,
\end{cases}
\qquad \alpha_-, \,\alpha_+ >0.
\end{equation}
When \(\alpha_-\ge \alpha_+\), the transformed distribution remains unimodal, and Eq.~\eqref{VPTUR} follows directly from the standard VP inequality. Thus, we focus on the non-unimodal case \(\alpha_-<\alpha_+\).

We set
\begin{equation}
\nu:=\frac{\mu}{\sigma},
\qquad
p:=\Pr(X\le 0).
\end{equation}
Since the transformation preserves the sign (\(X\le 0 \iff Y\le 0\)), we have
\begin{equation}
p=\Pr(Y\le 0)=\Phi(-\nu)
=1-\Phi(\nu).
\end{equation}
Here, $\Phi$ is the cumulative distribution function of the standard normal distribution.
In particular, the left-tail probability at the threshold zero is independent of \(\alpha_-\) and \(\alpha_+\).

Our goal is to prove the following Vysochanskij--Petunin-type bound:
\begin{equation}
\frac{\operatorname{Var}(X)}{\mathbb E[X]^2}
\ge
\begin{cases}
\dfrac{9p}{4-9p}, & 0<p\le \dfrac16,\\[2ex]
\dfrac{p+\frac13}{1-p}, & \dfrac16\le p\le \dfrac12.
\end{cases}
\label{eq:vp_goal_app}
\end{equation}

\paragraph{Reduction to the $\lambda\downarrow0$ case.}
The relative variance of \(X\) is scale-invariant. Hence, we may set \(\alpha_+=1\) and introduce the parameter
\begin{equation}
\lambda:=\frac{\alpha_-}{\alpha_+}\in(0,1).
\end{equation}
Then, $X$ is proportional to
\begin{equation}
X_\lambda = \lambda Y \mathbf{1}_{\{Y \le 0\}} + Y \mathbf{1}_{\{Y > 0\}}.
\end{equation}
The relative variance of $X_\lambda$ is given by
\begin{equation}
\rho(\lambda) 
= \frac{\mathbb E[X_\lambda^2]}{\mathbb E[X_\lambda]^2} - 1 
= \frac{\lambda^2 \mathbb E[Y^2 \mathbf{1}_{\{Y \le 0\}}] + \mathbb E[Y^2 \mathbf{1}_{\{Y > 0\}}]}{\bigl(\lambda \mathbb E[Y \mathbf{1}_{\{Y \le 0\}}] + \mathbb E[Y \mathbf{1}_{\{Y > 0\}}]\bigr)^2} - 1.
\end{equation}
We observe that the numerator $\mathbb E[X_\lambda^2]$ is strictly increasing in $\lambda$ since $\lambda \ge 0$ and $\mathbb E[Y^2 \mathbf{1}_{\{Y \le 0\}}] > 0$. Meanwhile, the expectation $\mathbb E[X_\lambda]$ is strictly decreasing in $\lambda$ because $\mathbb E[Y \mathbf{1}_{\{Y \le 0\}}] < 0$, but it remains strictly positive since $\mathbb E[X_\lambda] > \mathbb E[X_1] = \mu > 0$. 

Therefore, the ratio must be strictly increasing, thus for
\(
 0 < \lambda < 1,
\)
the smallest possible relative variance over this non-unimodal family is approached as \(\lambda\downarrow0\). Hence, it suffices to prove \eqref{eq:vp_goal_app} for the rectified variable \(X_0:=Y\mathbf 1_{\{Y>0\}}\).

We express the relative variance of $X_0$ in terms of the standard normal distribution.
Since
\begin{equation}
X_0
=
\begin{cases}
0, & \text{with probability } p=\Phi(-\nu),\\
Y\mid (Y>0), & \text{with probability } 1-p=\Phi(\nu),
\end{cases}
\end{equation}
we obtain
\(
\mathbb E[X_0] = \sigma\bigl(\nu \Phi(\nu)+\varphi(\nu)\bigr),
\)
and
\(
\mathbb E[X_0^2] = \sigma^2\bigl((1+\nu^2)\Phi(\nu)+\nu\varphi(\nu)\bigr),
\)
where \(\varphi\) is the probability density function of the standard normal distribution.
Here, we also denote
\begin{equation}
q := \frac{\varphi(\nu)}{\Phi(\nu)}.
\end{equation}
Let us define the squared coefficient of variation for the positive conditional part as
\begin{equation}
\operatorname{cv}^2(\nu)
:=
\frac{\operatorname{Var}(Y\mid Y>0)}{\mathbb E[Y\mid Y>0]^2}
=
\frac{1-\nu q-q^2}{(\nu+q)^2}.
\label{eq:cv_def_app}
\end{equation}
Then, the relative variance of \(X_0\), which corresponds to the infimum $\rho(0)$, can be written as
\begin{equation}
\rho(0)
=
\frac{\operatorname{Var}(X_0)}{\mathbb E[X_0]^2}
=
\frac{p+\operatorname{cv}^2(\nu)}{1-p}.
\label{eq:rho_rect_app}
\end{equation}
Thus, the desired VP-type lower bound follows if we prove
\begin{equation}
\operatorname{cv}^2(\nu) \ge
\begin{cases}
 \dfrac{5p}{4-9p} &\text{for } 0<p\le \dfrac16,\\
\dfrac13 &\text{for } \dfrac16\le p\le \dfrac12. 
\end{cases}
\end{equation}

\paragraph{(i) The branch $0<p\le 1/6$.}
The condition \(0< p\le 1/6\) is equivalent to \(5/6\le \Phi(\nu)< 1\), which implies \(\nu\ge \nu_0 := \Phi^{-1}(5/6)\).
For the proof of
\begin{equation}
\operatorname{cv}^2(\nu)\ge \frac{5p}{4-9p},
\end{equation}
we introduce the intermediate bound:
\begin{equation}
\frac{5p}{4-9p} \le \frac{1}{\nu^2+2} \le \operatorname{cv}^2(\nu).
\label{fns}    
\end{equation}
First, we derive the first part of Eq.~\eqref{fns} which is equivalent to \begin{equation}p\le \frac{4}{5\nu^2+19}.
\end{equation}
Defining
\begin{equation}
K(\nu) := 4-p(\nu)(5\nu^2+19),
\end{equation}
it is straightforward to show that \(K(\nu)>0\) for all \(\nu\ge\nu_0\). Since \(p'(\nu)=-\varphi(\nu)\), applying Mills' inequality~\cite{mills_ineq}: \(p(\nu)<\varphi(\nu)/\nu\) for \(\nu>0\) yields
\begin{equation}
K'(\nu)
=
\varphi(\nu)(5\nu^2+19)-10\nu p(\nu)
>
\varphi(\nu)(5\nu^2+9)>0.
\end{equation}
Thus, \(K\) is strictly increasing. Using \(p(\nu_0)=1/6\) and noting that \(\nu_0 < 1\) (since \(\Phi(1)\ge5/6\)), we have
\begin{equation}
K(\nu)\ge K(\nu_0) = 4-\frac{1}{6}(5\nu_0^2+19) = \frac{5}{6}(1-\nu_0^2) > 0.
\end{equation}
This proves the first inequality.

Next, we derive the second inequality of Eq.~\eqref{fns}, which simplifies to
\begin{equation}
\nu^3q+\nu^2q^2+4\nu q+3q^2 \le 2.
\label{eq:bridge_algebra_app}
\end{equation}
Since \(\Phi(\nu)\ge 5/6\) for \(\nu\ge\nu_0\), we have \(q=\varphi(\nu)/\Phi(\nu)\le {6}\varphi(\nu)/{5}\). Therefore, it suffices to prove
\begin{equation}
\frac65(\nu^3+4\nu)\varphi(\nu) + \frac{36}{25}(\nu^2+3)\varphi(\nu)^2 \le 2.
\end{equation}
We can readily verify this by evaluating the unique maxima of the functions \((\nu^3+4\nu)\varphi(\nu)\) and \((\nu^2+3)\varphi(\nu)^2\). Elementary calculus shows that
\begin{align*}
\frac65(\nu^3+4\nu)\varphi(\nu) + \frac{36}{25}(\nu^2+3)\varphi(\nu)^2
&<
\frac65(1.28) + \frac{36}{25}(0.25) \\
&= 1.896 < 2.
\end{align*}
This confirms Eq.~\eqref{eq:bridge_algebra_app}, and hence the second inequality holds.
Combining these two inequalities with Eq.~\eqref{eq:rho_rect_app} reduces to
\begin{equation}
\rho(0)
= \frac{p+\operatorname{cv}^2(\nu)}{1-p}
\ge \frac{p+\frac{5p}{4-9p}}{1-p}
= \frac{9p}{4-9p}.
\end{equation}

\paragraph{(ii) The branch $1/6\le p\le 1/2$.}

Since \(p=\Phi(-\nu)=1-\Phi(\nu)\), the condition \(1/6\le p\le 1/2\) is equivalent to \(1/2\le \Phi(\nu)\le 5/6\), which gives \(0\le \nu\le \Phi^{-1}(5/6)\). 
Since \(\Phi(1)>5/6\), it is sufficient to prove the desired estimate on the interval \(0\le \nu\le 1\).

The inequality \(\operatorname{cv}^2(\nu)\ge 1/3\) is equivalent to \(3\operatorname{cv}^2(\nu)-1 \ge 0\). Using \eqref{eq:cv_def_app}, the left-hand side evaluates to
\begin{equation}
3\operatorname{cv}^2(\nu)-1 = \frac{3-\nu^2-5\nu q-4q^2}{(\nu+q)^2}.
\label{eq:LHS}
\end{equation}
To verify the positivity of Eq.~\eqref{eq:LHS}, we use the standard fact that the left-tail inverse Mills' ratio \(q(\nu)\) is strictly convex on \([0,\infty)\)~\cite{Baricz2012MillsReciprocal,GASULL20141832}. Thus applying Jensen's inequality with the known values \(q(0)=\sqrt{2/\pi}<4/5\) and \(q(1)=\varphi(1)/\Phi(1)<3/10\), we obtain the upper bound
\begin{equation}
q(\nu)
\le
(1-\nu)\frac45+\nu\frac{3}{10}
=
\frac45-\frac{\nu}{2},
\qquad 0\le \nu\le 1.
\end{equation}
Since the numerator in Eq.~\eqref{eq:LHS} is decreasing as a function of
\(q\ge0\), the above upper bound gives
\begin{equation}
3-\nu^2-5\nu\left(\frac45-\frac{\nu}{2}\right)-4\left(\frac45-\frac{\nu}{2}\right)^2
=
\frac{25\nu^2-40\nu+22}{50}.
\end{equation}
This quadratic is strictly positive for all \(\nu\). Hence, \(\operatorname{cv}^2(\nu)\ge 1/3\) holds, and by \eqref{eq:rho_rect_app}, we get
\begin{equation}
\rho(0) \ge \frac{p+\frac13}{1-p}, \qquad \frac16\le p\le \frac12.
\end{equation}

\paragraph{Conclusion.}
Combining branches (i) and (ii), we obtain
\begin{equation}
\rho(0) \ge
\begin{cases}
\dfrac{9p}{4-9p}, & 0<p\le \dfrac16,\\[2ex]
\dfrac{p+\frac13}{1-p}, & \dfrac16\le p\le \dfrac12.
\end{cases}
\end{equation}
Finally, since \(\rho(\lambda)\ge\rho(0)\) by the monotonicity established earlier, the same bound holds for every \(0<\lambda<1\), or equivalently for every \(0<\alpha_-<\alpha_+\). Hence,
\begin{equation*}
\frac{\operatorname{Var}(X)}{\mathbb E[X]^2}
\ge
\begin{cases}
\dfrac{9\Pr(X\le 0)}{4-9\Pr(X\le 0)}, & 0\leq\Pr(X\le 0)\le \dfrac16,\\[2ex]
\dfrac{\Pr(X\le 0)+\frac13}{1-\Pr(X\le 0)}, & \dfrac16\le \Pr(X\le 0)\leq\dfrac12.
\end{cases}
\end{equation*}

\clearpage
\bibliography{reference}

\end{document}

%% file: setting.tex
\usepackage{ifthen}

\newboolean{showcomment}
\setboolean{showcomment}{true}

\usepackage[whole]{bxcjkjatype}
\usepackage[normalem]{ulem}
\usepackage{soul}
\definecolor{electricindigo}{rgb}{0.44, 0.0, 1.0}
\definecolor{electricpurple}{rgb}{0.75, 0.0, 1.0}
\definecolor{electricultramarine}{rgb}{0.25, 0.0, 1.0}
\definecolor{americanrose}{rgb}{1.0, 0.01, 0.24}
\ifthenelse{\boolean{showcomment}}{
    \newcommand{\hase}[1]{\textcolor{americanrose}{{\bf 長谷川: }#1}}
    \newcommand{\indo}[1]{\textcolor{electricpurple}{{\bf 犬童: }#1}}
}{
    \newcommand{\hase}[1]{}
    \newcommand{\indo}[1]{}
}

\definecolor{coquelicot}{rgb}{1.0, 0.22, 0.0}
\definecolor{shamrockgreen}{rgb}{0.0, 0.62, 0.38}
\definecolor{tangelo}{rgb}{0.98, 0.3, 0.0}
\definecolor{royalblue(web)}{rgb}{0.25, 0.41, 0.88}
\definecolor{ceruleanblue}{rgb}{0.16, 0.32, 0.75}
\definecolor{carminered}{rgb}{1.0, 0.0, 0.22}
\ifthenelse{\boolean{showcomment}}{

}{

}

\usepackage{lineno}

%% file: graph7.pdf_tex
\begingroup%
  \makeatletter%
  \providecommand\color[2][]{%
    \errmessage{(Inkscape) Color is used for the text in Inkscape, but the package 'color.sty' is not loaded}%
    \renewcommand\color[2][]{}%
  }%
  \providecommand\transparent[1]{%
    \errmessage{(Inkscape) Transparency is used (non-zero) for the text in Inkscape, but the package 'transparent.sty' is not loaded}%
    \renewcommand\transparent[1]{}%
  }%
  \providecommand\rotatebox[2]{#2}%
  \newcommand*\fsize{\dimexpr\f@size pt\relax}%
  \newcommand*\lineheight[1]{\fontsize{\fsize}{#1\fsize}\selectfont}%
  \ifx\svgwidth\undefined%
    \setlength{\unitlength}{129.42142156bp}%
    \ifx\svgscale\undefined%
      \relax%
    \else%
      \setlength{\unitlength}{\unitlength * \real{\svgscale}}%
    \fi%
  \else%
    \setlength{\unitlength}{\svgwidth}%
  \fi%
  \global\let\svgwidth\undefined%
  \global\let\svgscale\undefined%
  \makeatother%
  \begin{picture}(1,0.83781233)%
    \lineheight{1}%
    \setlength\tabcolsep{0pt}%
    \put(0,0){\includegraphics[width=\unitlength,page=1]{graph7.pdf}}%
    \put(0.11039335,0.21912028){\color[rgb]{0,0,0}\makebox(0,0)[lt]{\lineheight{1.25}\smash{\begin{tabular}[t]{l}{\large$\Gamma_r^-$}\\\end{tabular}}}}%
    \put(0.14198779,0.42866762){\color[rgb]{0,0,0}\makebox(0,0)[lt]{\lineheight{1.25}\smash{\begin{tabular}[t]{l}{\large$-\sigma$}\\\end{tabular}}}}%
    \put(0.80177616,0.66418984){\color[rgb]{0,0,0}\makebox(0,0)[lt]{\lineheight{1.25}\smash{\begin{tabular}[t]{l}{\large$\Gamma_r^+$}\\\end{tabular}}}}%
    \put(0,0){\includegraphics[width=\unitlength,page=2]{graph7.pdf}}%
    \put(0.71980383,0.29700898){\color[rgb]{0,0,0}\makebox(0,0)[lt]{\lineheight{1.25}\smash{\begin{tabular}[t]{l}{\large$\sigma$}\\\end{tabular}}}}%
    \put(0,0){\includegraphics[width=\unitlength,page=3]{graph7.pdf}}%
    \put(0.3840774,0.59642924){\color[rgb]{0,0,0}\makebox(0,0)[lt]{\lineheight{1.25}\smash{\begin{tabular}[t]{l}{\large$\phi$}\end{tabular}}}}%
    \put(0.51775948,0.24194566){\color[rgb]{0,0,0}\makebox(0,0)[lt]{\lineheight{1.25}\smash{\begin{tabular}[t]{l}{\large$\phi^\dagger$}\end{tabular}}}}%
    \put(0,0){\includegraphics[width=\unitlength,page=4]{graph7.pdf}}%
  \end{picture}%
\endgroup%

%% file: drawing02.pdf_tex
\begingroup%
  \makeatletter%
  \providecommand\color[2][]{%
    \errmessage{(Inkscape) Color is used for the text in Inkscape, but the package 'color.sty' is not loaded}%
    \renewcommand\color[2][]{}%
  }%
  \providecommand\transparent[1]{%
    \errmessage{(Inkscape) Transparency is used (non-zero) for the text in Inkscape, but the package 'transparent.sty' is not loaded}%
    \renewcommand\transparent[1]{}%
  }%
  \providecommand\rotatebox[2]{#2}%
  \newcommand*\fsize{\dimexpr\f@size pt\relax}%
  \newcommand*\lineheight[1]{\fontsize{\fsize}{#1\fsize}\selectfont}%
  \ifx\svgwidth\undefined%
    \setlength{\unitlength}{411.3573642bp}%
    \ifx\svgscale\undefined%
      \relax%
    \else%
      \setlength{\unitlength}{\unitlength * \real{\svgscale}}%
    \fi%
  \else%
    \setlength{\unitlength}{\svgwidth}%
  \fi%
  \global\let\svgwidth\undefined%
  \global\let\svgscale\undefined%
  \makeatother%
  \begin{picture}(1,0.7993307)%
    \lineheight{1}%
    \setlength\tabcolsep{0pt}%
    \put(0,0){\includegraphics[width=\unitlength,page=1]{drawing02.pdf}}%
    \put(0.7722496,0.59854634){\color[rgb]{1,1,1}\makebox(0,0)[lt]{\lineheight{1.25}\smash{\begin{tabular}[t]{l}{\large $\beta_2$}\end{tabular}}}}%
    \put(0.73914249,0.14634366){\color[rgb]{1,1,1}\makebox(0,0)[lt]{\lineheight{1.25}\smash{\begin{tabular}[t]{l}{\large $\beta_1$}\end{tabular}}}}%
    \put(0.07728286,0.39977289){\color[rgb]{1,1,1}\makebox(0,0)[lt]{\lineheight{1.25}\smash{\begin{tabular}[t]{l}{\large $\beta_3$}\end{tabular}}}}%
    \put(0.67743248,0.41337613){\color[rgb]{0,0,0}\makebox(0,0)[lt]{\lineheight{1.25}\smash{\begin{tabular}[t]{l}{\Large $\ket{e_A}$}\end{tabular}}}}%
    \put(0.15627858,0.08545811){\color[rgb]{0,0,0}\makebox(0,0)[lt]{\lineheight{1.25}\smash{\begin{tabular}[t]{l}{\Large $\ket{g}$}\\\end{tabular}}}}%
    \put(0.24781321,0.64884717){\color[rgb]{0,0,0}\makebox(0,0)[lt]{\lineheight{1.25}\smash{\begin{tabular}[t]{l}{\Large $\ket{e_B}$}\end{tabular}}}}%
  \end{picture}%
\endgroup%

%% file: drawing3.pdf_tex
\begingroup%
  \makeatletter%
  \providecommand\color[2][]{%
    \errmessage{(Inkscape) Color is used for the text in Inkscape, but the package 'color.sty' is not loaded}%
    \renewcommand\color[2][]{}%
  }%
  \providecommand\transparent[1]{%
    \errmessage{(Inkscape) Transparency is used (non-zero) for the text in Inkscape, but the package 'transparent.sty' is not loaded}%
    \renewcommand\transparent[1]{}%
  }%
  \providecommand\rotatebox[2]{#2}%
  \newcommand*\fsize{\dimexpr\f@size pt\relax}%
  \newcommand*\lineheight[1]{\fontsize{\fsize}{#1\fsize}\selectfont}%
  \ifx\svgwidth\undefined%
    \setlength{\unitlength}{841.88976378bp}%
    \ifx\svgscale\undefined%
      \relax%
    \else%
      \setlength{\unitlength}{\unitlength * \real{\svgscale}}%
    \fi%
  \else%
    \setlength{\unitlength}{\svgwidth}%
  \fi%
  \global\let\svgwidth\undefined%
  \global\let\svgscale\undefined%
  \makeatother%
  \begin{picture}(1,0.38720539)%
    \lineheight{1}%
    \setlength\tabcolsep{0pt}%
    \put(0,0){\includegraphics[width=\unitlength,page=1]{drawing3.pdf}}%
    \put(0.25370774,0.17417255){\color[rgb]{0,0,0}\makebox(0,0)[lt]{\lineheight{1.25}\smash{\begin{tabular}[t]{l}$x_1$\end{tabular}}}}%
    \put(0.40202551,0.17417255){\color[rgb]{0,0,0}\transparent{0.97435898}\makebox(0,0)[lt]{\lineheight{1.25}\smash{\begin{tabular}[t]{l}$x_2$\end{tabular}}}}%
    \put(0.70399096,0.17417255){\color[rgb]{0,0,0}\transparent{0.97435898}\makebox(0,0)[lt]{\lineheight{1.25}\smash{\begin{tabular}[t]{l}$x_N$\end{tabular}}}}%
    \put(0.5324603,0.18993648){\color[rgb]{0,0,0}\transparent{0.97435898}\makebox(0,0)[lt]{\lineheight{1.25}\smash{\begin{tabular}[t]{l}. . .\end{tabular}}}}%
    \put(0,0){\includegraphics[width=\unitlength,page=2]{drawing3.pdf}}%
    \put(0.31054175,0.02100875){\color[rgb]{0,0,0}\makebox(0,0)[lt]{\lineheight{1.25}\smash{\begin{tabular}[t]{l}$dQ_1$\end{tabular}}}}%
    \put(0.81441795,0.02100875){\color[rgb]{0,0,0}\makebox(0,0)[lt]{\lineheight{1.25}\smash{\begin{tabular}[t]{l}$dQ_N$\end{tabular}}}}%
    \put(0.48554957,0.02100875){\color[rgb]{0,0,0}\makebox(0,0)[lt]{\lineheight{1.25}\smash{\begin{tabular}[t]{l}$dQ_2$\end{tabular}}}}%
    \put(0.11630212,0.02100875){\color[rgb]{0,0,0}\makebox(0,0)[lt]{\lineheight{1.25}\smash{\begin{tabular}[t]{l}$dQ_0$\end{tabular}}}}%
    \put(0,0){\includegraphics[width=\unitlength,page=3]{drawing3.pdf}}%
  \end{picture}%
\endgroup%